\documentclass[journal=jpcbfk,manuscript=article]{achemso}

\usepackage[version=3]{mhchem} 
\usepackage{textcomp}

\usepackage{color}
\newcommand{\PE}{\textit{n-}C\textsubscript{720}H\textsubscript{1442}~}
\newcommand{\ie}{i.e.,~}
\newcommand{\eg}{e.g.,~}

\author{Kyle Wm. Hall}
\affiliation[Temple Chemistry]
{Department of Chemistry, Temple University, Philadelphia, Pennsylvania 19122, USA}
\alsoaffiliation[ICMS]
{Institute for Computational Molecular Science, Philadelphia, Pennsylvania 19122, USA}
\email{k.wm.hall@temple.edu}
\author{Simona Percec}
\affiliation[Temple Chemistry]
{Department of Chemistry, Temple University, Philadelphia, Pennsylvania 19122, USA}
\author{Wataru Shinoda}
\affiliation[Nagoya Chemistry]
{Department of Materials Chemistry, Nagoya University, Furo-cho, Chikusa-ku, Nagoya 464-8603, Japan}
\author{Michael L. Klein}
\affiliation[Temple Chemistry]
{Department of Chemistry, Temple University, Philadelphia, Pennsylvania 19122, USA}
\alsoaffiliation[ICMS]
{Institute for Computational Molecular Science, Philadelphia, Pennsylvania 19122, USA}

\title[Property Decoupling across the Nucleus-Melt Interface during Polymer Crystal Nucleation]
  {Property Decoupling across the Nucleus-Melt Interface during Polymer Crystal Nucleation}


\begin{document}

\makeatletter
\setlength\acs@tocentry@height{8.25cm}
\setlength\acs@tocentry@width{4.45cm}
\makeatother
\begin{tocentry}
\begin{center}
\includegraphics[width=8.25cm]{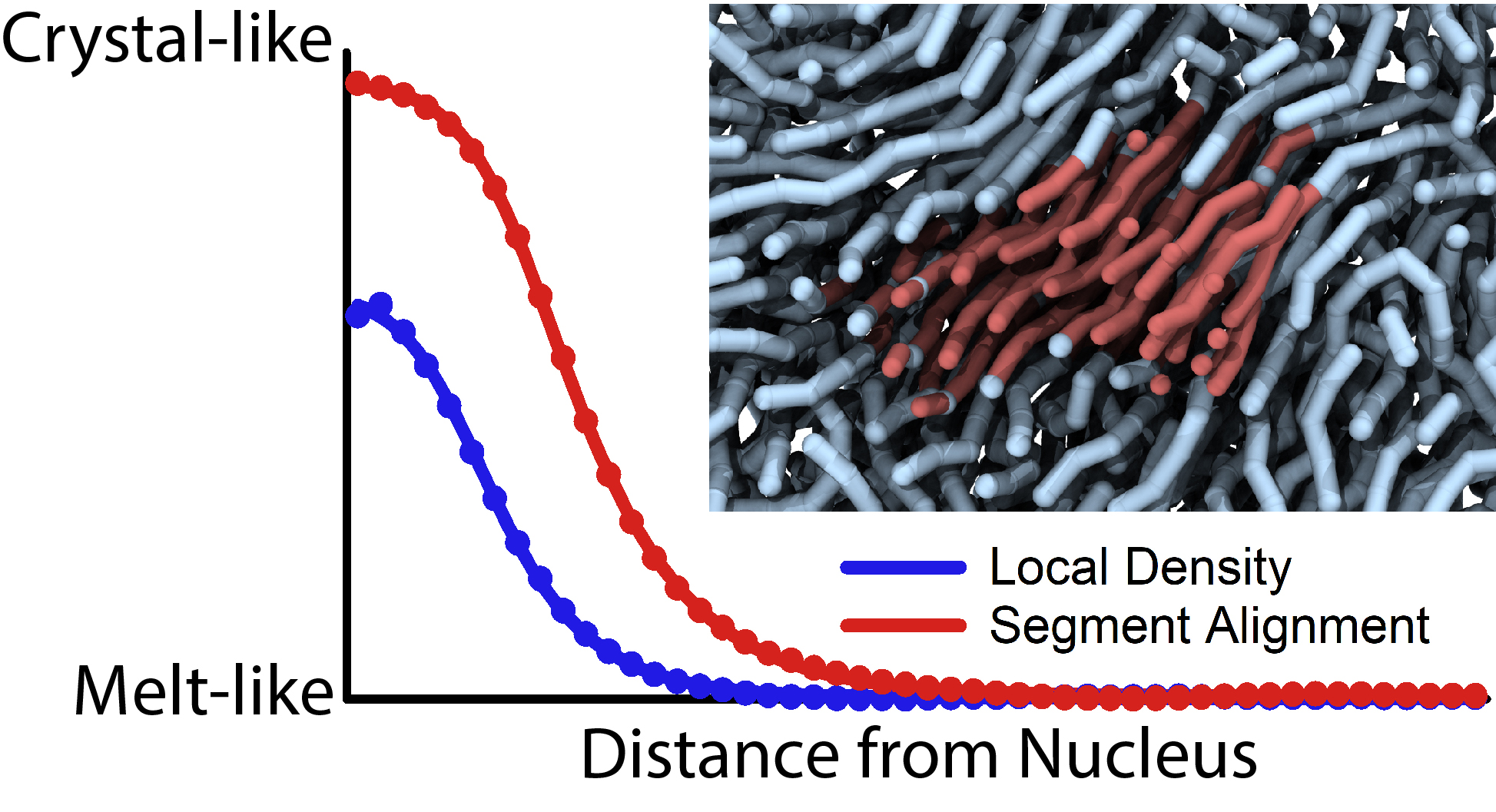}
\end{center}
\end{tocentry}

\begin{abstract}
Spatial distributions are presented that quantitatively capture how polymer properties (e.g., segment alignment, density, and potential energy) vary with distance from nascent polymer crystals (nuclei) in prototypical polyethylene melts. It is revealed that the spatial extent of nuclei and their interfaces is metric-dependent as is the extent to which nucleus interiors are solid-like. As distance from a nucleus increases, some properties, such as density, decay to melt-like behavior more rapidly than polymer segment alignment, indicating that a polymer nucleus resides in a nematic-like droplet. This nematic-like droplet region coincides with enhanced formation of ordered polymer segments that are not part of the nucleus. It is more favourable to find non-constituent ordered polymer segments near a nucleus than in the surrounding metastable melt, pointing to the possibility of one nucleus inducing the formation of other nuclei. These findings provide a conceptual bridge between polymer crystal nucleation under non-flow and flow conditions, and are used to rationalize previous results.
\end{abstract}

\section{Introduction}
Many polymers (\eg polyethylene, polyamides, polyesters, etc.) are semi-crystalline; their solid-states contain both crystalline and amorphous domains. The characteristics of these domains are governed by phenomena that occur during polymer processing, dictating the properties of polymeric materials and their potential uses (e.g., see ref.~\citenum{ParkRutledge2018,ShenNatureNano2010}). As such, understanding and controlling polymer crystallization is of great scientific and technological importance, especially in light of the ubiquity of polymeric materials.  However, scientific understanding of polymer crystallization remains incomplete and an active area of research as evidenced by a number of recent reviews.\cite{CuiEtAlChemRev2018,YuePolyCrystal2018,LotzMacro2017,SchickJPhysCond2017,ZhangCrystals2017} In the intervening decades since Keller's seminal work\cite{KellerDFS1958,KellerPhilMag1957} revealing that polyethylene chains adopt folded conformations upon crystallization, there has been a large number of experimental,\cite{LiuTianMacro2014,KanayaMacro2013,ZhaoMacro2013,GutierrezEtAlMacro2004,SekiMacro2002,BalzanoPRL2008,WangEtAlMacro2000,RyanFarad1999,SamonMacro1999,TerrillEtAlPolymer1998} theoretical\cite{SadlerNat1987,SadlerGilmerPRL1986,LauritzenJAppPhys1973,PointMacro1979,PointFaraday1979,FrankTosi1961,PriceJCP1961,Lauritzen1960} and computational\cite{ZhaiEtAlMacro2019,YamamotoMacro2019,HagitaEtAlJCP2019,LiEtAlJCP2019,MoyassariPolymer2019,MoyassariEtAlMacro2019,HuEtAlPolymer2019,ZhangLarsonJCP2019,AnwarGrahamJCP2019,HallPolymers2020,HallShapeJCP2019,HallHeatJCP2019,ZhangLarsonMacro2018,HagitaAIPAdvances2018,VerhoMacro2018,SliozbergMacro2018,NieEtAlMacro2018,KumarMacro2017,BourqueEtAlJPCB2017,BourqueMacro2016,LuoPolymer2017,LuoEtAlMacro2016,TangetalJCP2018,TangPRM2017,WelchJCP2017,WangEtAlCTC2015,LuoSommerPRL2014,AnwarEtAlJCP2014,YamamotoJCP2013,YiMacro2013,AnwarJCP2013,LuoSommerMacro2011,YamamotoJCP2010,YamamotoJCP2008,HuFrenkelMacro2004,HuEtAlMacro2002,MeyerJCP2001,MuthukumarWelchPolymer,DoyePolymer2000,DoyeFrenkelJCP19992,DoyeFrenkelJCP19991,LiuMuthukumar1998} studies probing polymer crystallization and related phenomenology.

Crystal nucleation corresponds to the very earliest stages of crystallization in which a nascent crystal (i.e., nucleus) emerges from a metastable melt or solution. Crystal nucleation is being increasingly studied using simulations as they provide direct access to molecular details at high spatiotemporal resolutions, something challenging to achieve experimentally. Previous \textit{in silico} work has probed polymer crystal nucleation in the context of isolated chains,\cite{HagitaAIPAdvances2018,MuthukumarWelchPolymer,LiuMuthukumar1998} solutions,\cite{LuoEtAlMacro2016,HuFrenkelMacro2004,HuEtAlMacro2002,LiuMuthukumar1998} and melts.\cite{MoyassariEtAlMacro2019,ZhangLarsonJCP2019,ZhangLarsonMacro2018,AnwarGrahamJCP2019,HallPolymers2020,HallShapeJCP2019,HallHeatJCP2019,WelchJCP2017,WangEtAlCTC2015,AnwarEtAlJCP2014,YiMacro2013,YamamotoJCP2010,MeyerJCP2001} Simulations have also been used to elucidate the effects of molecular weight distribution,\cite{ZhaiEtAlMacro2019,LiEtAlJCP2019,MoyassariPolymer2019,MoyassariEtAlMacro2019,NieEtAlMacro2018} chain topology (\eg linear vs. ring chains),\cite{HagitaEtAlJCP2019} chain branching,\cite{MoyassariEtAlMacro2019,HuEtAlPolymer2019,ZhangLarsonMacro2018} and cross-linking\cite{PaajanenPolymer2019} on polymer crystal nucleation. There has been much interest in the evolution of chain conformations,\cite{MoyassariPolymer2019,MoyassariEtAlMacro2019,HallHeatJCP2019,WangEtAlCTC2015,YamamotoJCP2010,YamamotoJCP2008,LiuMuthukumar1998} topological details related to chain folding,\cite{ZhaiEtAlMacro2019,YamamotoMacro2019,MoyassariEtAlMacro2019,MoyassariPolymer2019,HallPolymers2020,HallShapeJCP2019,HallHeatJCP2019,MorthomasEtAlPhysRevE2017,LuoEtAlMacro2016,YiMacro2013,MeyerJCP2001} and connecting entanglements/disentanglement to observed crystallization phenomenology.\cite{ZhaiEtAlNano,LuoPolymer2017,LuoEtAlMacro2016,LuoSommerPRL2014} For example, Luo and Sommer\cite{LuoSommerPRL2014} demonstrated that local entanglements play a key role in polymer crystallization memory effects, and revealed connections between local entanglement lengths, stem lengths and densities. Through simulations, it has also been demonstrated that quenching polymer melts below their nematic-isotropic transition temperatures can enhance crystallization via the formation of metastable nematic precursors.\cite{ZhangLarsonJCP2019} Other recent studies have probed flow-induced crystal nucleation in polymer melts\cite{AnwarGrahamJCP2019,YamamotoMacro2019,SliozbergMacro2018,NieEtAlMacro2018,AnwarEtAlJCP2014} and short-chain alkane systems.\cite{NicholsonRutledgeJCP2016}
Additional \textit{in silico} work\cite{LuoPolymer2017,YamamotoJCP2013} has explored crystal nucleation in supported and free-standing thin films, probing the effects of polymer-melt and polymer-substrate interfaces on nucleation phenomenology. While there has been much work on polymer crystal nucleation, the interfacial structuring of nuclei remains underexplored. Structural descriptions of polymer nucleus-melt interfaces are generally qualitative and focused on topological details. For example, previous simulation-based work\cite{SliozbergMacro2018} qualitatively revealed that chain ends are preferentially partitioned to crystal-melt interfaces, and that chain entanglements are generally relegated to amorphous domains during the crystallization of entangled polymer melts under uniaxial strain. A quantitative, spatially-resolved picture of the nucleus-melt interfacial region has yet to emerge. 

\begin{figure}[!b]
\centering
\includegraphics[width=0.5\columnwidth]{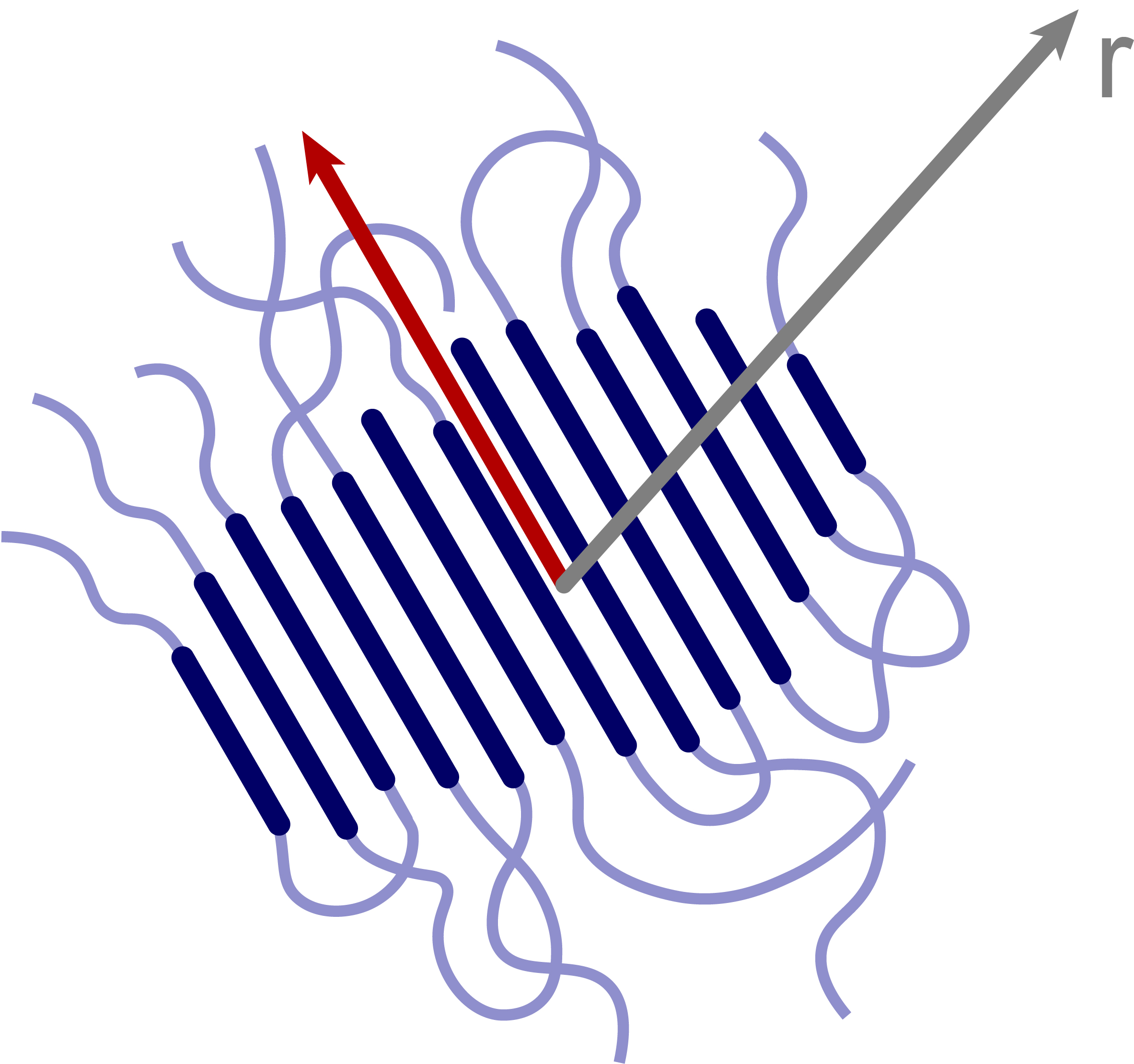}
\caption{A schematic of a polymer nucleus. The dark blue lines correspond to the ordered crystalline polymer chain segments that form the nucleus (i.e., stems).  The red arrow indicates the direction of the stems (i.e., the nematic director of the nucleus). The pale blue lines indicate polymer chain segments that are amorphous and not part of the nucleus. A fold corresponds to an amorphous segment that connects two stems that are part of the same nucleus. The surrounding melt has not been shown for visual clarity. The work reported herein focuses on the evolution of average segment properties with distance from the center of mass of a nucleus as represented by the gray arrow.}
\label{fig:schematic}
\end{figure}

To address this knowledge gap, we have conducted molecular dynamics (MD) simulations of crystal nucleation in entangled polyethylene melts, and thereby provide a quantitative understanding of variations in polymer properties in the vicinity of nuclei. This study explores how polymer properties vary from the center of a nucleus into the surrounding metastable melt (see the schematic in Fig.~\ref{fig:schematic}), revealing that there is a partial spatial decoupling of segment properties at the nucleus interface. In particular, nuclei arising during quiescent, non-flow crystallization apparently reside in nematic droplets, which has broad implications for crystallization phenomenology as well as the interpretation of both experimental and computational results.

\section{Methods}
\subsection{Crystallization Simulations}
\begin{figure}[!ht]
\centering
\includegraphics[width=1.0\columnwidth]{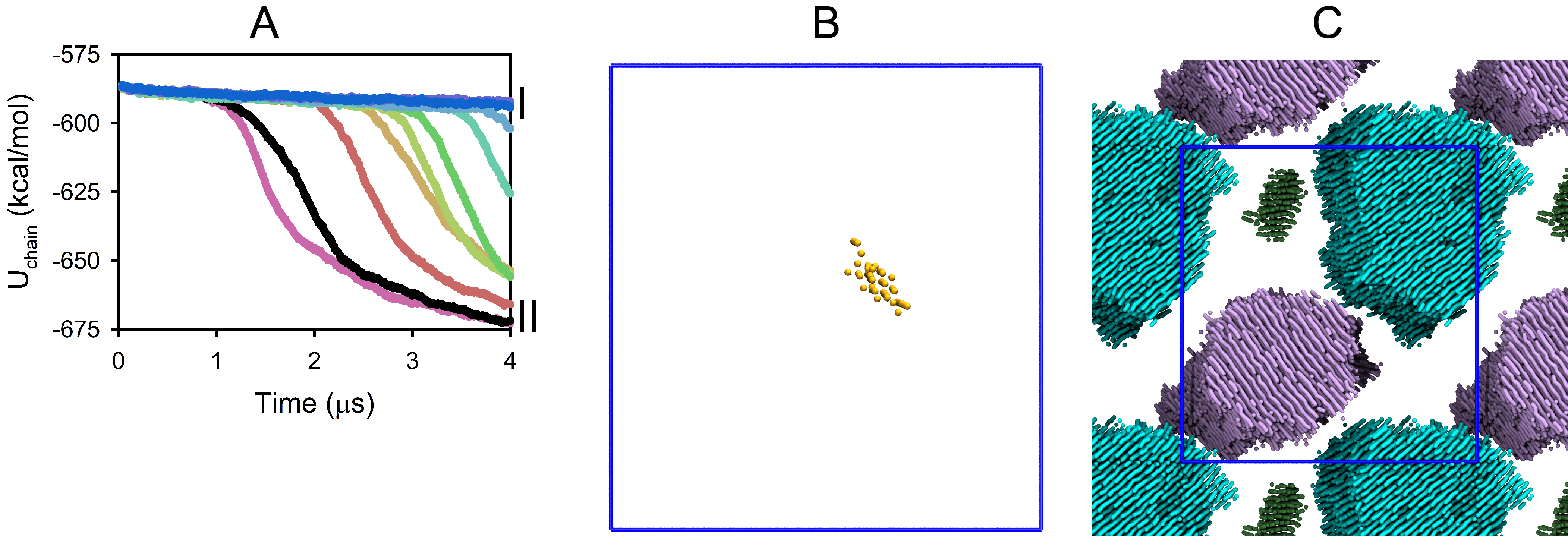}
\caption{Temporal evolution of the nucleation simulations and their final states. A) The average molecular potential energy ($U_{chain}$) curves for this study's ten nucleation simulations at 285 K. Crystallization is associated with a large decrease in $U_{chain}$. B) A zoomed-in view of the final configuration from the simulation corresponding to the dark blue curve labelled I in Panel A. Only the ordered polymer chain segments composing the largest nucleus are shown for visual clarity. The nucleus corresponds to approximately 100 carbon atom. According to previous work,\cite{HallShapeJCP2019} the critical nucleus size for the conditions and model considered in this study is $\sim$600 carbon atoms. The blue box corresponds to the simulation cell. C) A zoomed-out view of the final configuration from the simulation corresponding to the black curve labelled II in Panel A. The simulation cell corresponds to the blue box; all simulations were performed under periodic boundary conditions for this study. All nuclei larger than the critical nucleus size are shown, and each nucleus is visualized using a different color. The methodological details of how nuclei were extracted from simulation snapshots are provided in the main text.}
\label{fig:potEner}
\end{figure}

The MD simulation details for this study are the same as for our earlier work;\cite{HallShapeJCP2019} see ref.~\citenum{HallShapeJCP2019} for full details. Briefly, ten high-temperature, entangled \PE melts were prepared, and then sequentially cooled to crystal forming conditions (285 K) where they were simulated for $\sim$4 $\mu$s. The simulations probed polymer crsytal nucleation under quiescent, non-flow conditions. Crystal nucleation took place on the microsecond timescale,
 and two melts did not crystallize during the 4-$\mu$s simulation window (see Fig.~\ref{fig:potEner}). During each simulation, system configurations (snapshots) were saved every 20,000 iterations ($\sim$0.4 ns). The coarse-grain Shinoda-DeVane-Klein (SDK) model\cite{ShinodaMolSim2007} was used to represent \PE. More specifically, \textendash(CH\textsubscript{2})\textsubscript{3}\textendash~and \textendash{CH\textsubscript{2}}CH\textsubscript{2}CH\textsubscript{3} segments along the polymer chain backbones were represented using the CM and CT coarse-grain beads.\cite{ShinodaMolSim2007} As such, the terms ``bead'' and ``segment'' are used interchangeably throughout this study. We have previously demonstrated that the SDK model is an appropriate coarse-grain model for simulating polyethylene systems and processes.\cite{HallModelJCP2019} The simulations were conducted using the Large-scale Atomic/Molecular Massively Parallel Simulator (LAMMPS).\cite{Plimpton1995}

\subsection{Nucleus Extraction}
Nuclei were extracted from each snapshot using the same approach as our previous work,\cite{HallShapeJCP2019,HallModelJCP2019,HallHeatJCP2019} an approach based on the $P_2$ order parameter. The $P_2$ order parameter captures the degree to which a polymer segment is locally aligned with its neighbouring segments. The $P_2$ order parameter and variations thereof have been used to study polymer crystallization by many other researcher groups (\eg see ref.~\citenum{MoyassariEtAlMacro2019,YamamotoMacro2019,HuEtAlPolymer2019,ZhangLarsonJCP2019,SliozbergMacro2018,NicholsonRutledgeJCP2016,BourqueMacro2016,YiMacro2013,YamamotoJCP2013,LiuMuthukumar1998}). For the purposes of this study, the backbone direction at a given coarse-grain segment (bead $i$) was estimated using the vector connecting its intramolecular nearest neighbours (\ie beads $i-1$ and $i+1$). The exception to this was the segments at the ends of the polymer chains, which only have one intramolecular nearest neighbour. In this case, the backbone direction was estimated using the vector connecting each terminal segment to its nearest intramolecular neighbour. In turn, the $P_2$ order parameter value for the i\textsuperscript{th} polymer segment was calculated according to: 
\begin{equation}
P_2=\bigg \langle \frac{3cos^{2}\theta_{ij}-1}{2} \bigg \rangle
\label{eq:p2}
\end{equation}
where $\theta_{ij}$ is the angle between the backbone direction at segment $i$ and the backbone direction at neighbouring segment $j$. The angular brackets in Eq.~\ref{eq:p2} indicate local averaging over all intramolecular and intermolecular neighbours within 0.635 nm of segment $i$. For reference, average $P_2$ values of 1.0, 0 and -0.5 indicate that a polymer segment is perfectly aligned, randomly oriented, and perpendicular with respect to its neighbouring polymer chain segments. All beads with $P_2 \geq 0.85$ were labelled ordered (\ie crystalline).\cite{HallHeatJCP2019} The above protocol and cutoffs were established as part of our previous work\cite{HallHeatJCP2019} probing polyethylene crystallization with the SDK model. The polymer segments in each simulation snapshot were thus labelled ordered or disordered (\ie crystalline or non-crystalline). Cluster analysis was then performed on the ordered polymer segments in each snapshot to extract its nuclei; ordered segments within 0.635 nm of each other were taken to be part of the same cluster.\cite{HallHeatJCP2019}

\subsection{Local Properties}
The local properties of the segments (beads) in each snapshot were also assessed in order to construct spatial property distributions of nuclei and their surroundings. The following four properties were measured in addition to the $P_2$ order parameter. \newline \newline
\textbf{Potential Energy (U):} The potential energy of each segment was estimated as the sum of the potential energy contributions arising from its bonded and non-bonded interactions as specified by the SDK model.\cite{ShinodaMolSim2007}
\newline \newline
\textbf{Density ($\rho$):} A Voronoi tessellation was performed on each snapshot according to the positions of its constituent polymer segments (beads). For reference, a Voronoi tessellation partitions the simulation cell into a set of sub-cells, specifically one sub-cell per segment, such that the sub-cell associated with a given polymer segment corresponds to the region of space for which that segment is the nearest segment. In turn, the local density associated with each segment was estimated by dividing the mass of the segment by its Voronoi sub-cell volume.
\newline \newline
\textbf{Nucleus-Segment Alignment ($S$):}
The nematic director of a nucleus corresponds to the overall direction of its constituent ordered polymer chain segments, and is schematically represented by the red arrow in Fig.~\ref{fig:schematic}. The nematic director of each nucleus was extracted in accordance with the procedure of Eppanga and Frenkel.\cite{EppangeMolPhys1984} The tensor order parameter $Q$ was constructed based on the backbone vectors of the segments comprising a nucleus, and then the nematic director of that nucleus was taken to be the eigenvector associated with the largest eigenvalue of $Q$ (see ref.~\citenum{EppangeMolPhys1984}).

$S$ quantifies the alignment of polymer segments with the nematic director of a particular nucleus, and was calculated according to:
\begin{equation}
S=\frac{3cos^{2}\theta_{i}-1}{2}
\label{eq:S}
\end{equation}
where $\theta_{i}$ is the angle between the nematic director of the specified nucleus and the polymer backbone direction at segment $i$; the latter was estimated using the same procedure as for the $P_2$ calculations. As with the $P_2$ order parameter, average $S$ values of 1.0, 0 and -0.5 indicate perfect alignment, random orientations, and perpendicular configurations, respectively. However, $S$ probes nucleus-segment alignment whereas $P_2$ probes local segment-segment alignment.
\newline \newline
\textbf{Steinhardt-based Order ($q_{6}q_{6}^{*}$):}
The $q_{6}q_{6}^{*}$ metric was recently introduced by Zhang and Larson\cite{ZhangLarsonJCP2019,ZhangLarsonMacro2018} to probe polymer crystallization. It is based on the $q_{6m}$ metric as introduced by Steinhardt and coworkers,\cite{SteinhardtPRC1983} and the work of Auer and Frenkel.\cite{AuerFrenkelJCP2004} In this study, $q_{6m}$ was calculated for a given segment ($i$) according to:
\begin{equation}
q_{6m}(i)=\langle Y_{6m}(\theta(\vec{r}_{ij}),\phi(\vec{r}_{ij})\rangle
\label{eq:q6}
\end{equation}
where $Y_{6m}$ corresponds to the $m$ component of the degree-6 spherical harmonic (\ie l=6), $\vec{r}_{ij}$ is the vector from segment $i$ to a neighbouring segment ($j$), and $\theta$ and $\phi$ indicate the spherical polar orientation of $\vec{r}_{ij}$ in a fixed frame of reference centered on segment $i$. The angular brackets indicate averaging over all neighbouring segments within 0.635 nm of segment $i$. In turn, the $q_{6}q_{6}^{*}$ value of a segment was determined according to:
\begin{equation}
q_{6}q_{6}^{*}=\sum_{j=1}^{N_{i}} \sum_{m=-6}^{m=6}  q_{6m}(i) q_{6m}^{*}(j)
\label{eq:q6q6}
\end{equation}
where the first sum is over all neighbouring segments within 0.635 nm of segment $i$ excluding itself, and $q_{6m}^{*}(j)$ is the complex conjugate of $q_{6m}(j)$. Given that $q_{6}q_{6}^{*}$ depends on the $q_{6m}$ values of segments within 0.635 nm~of segment $i$, and that the $q_{6m}$ values of these neighboring segments depend on beads within 0.635 nm of them, the $q_{6}q_{6}^{*}$ value of a segment depends indirectly on segments up to $\sim$1.27 nm away from that segment. In turn, $q_{6}q_{6}^{*}$
probes an approximately eightfold larger region of space than the $P_2$ order parameter. Note that a distance cutoff of 0.635 nm was used for Eq.~\ref{eq:q6} and~\ref{eq:q6q6}  instead of the 0.54-nm cutoff as in Zhang and Larson's work\cite{ZhangLarsonJCP2019,ZhangLarsonMacro2018} because this study uses a different molecular model. Zhang and Larson used atomistic models whereas this study uses the coarse-grain SDK model.\cite{ShinodaMolSim2007}

 \subsection{Spatial Property Distributions}
 The property values and nuclei extracted from each snapshot were in turn used to construct spatial distributions capturing the evolution of segment properties in the vicinity of nuclei corresponding to 240-360 carbon atoms inclusive. This size range was chosen as it allowed for good sampling (\ie thousands of nuclei) while still being comparable to the critical nucleus size for the polyethylene melts considered in this study (\ie $\sim$600 carbon atoms based on our previous work\cite{HallShapeJCP2019}).
 
 A principle axis system was determined for each extracted nucleus by calculating the three eigenvectors and eigenvalues of its radius of gyration tensor. The eigenvalues indicate the spatial extent of the particles along each eigenvector (\ie axis). As previously demonstrated,\cite{HallShapeJCP2019}  a pre-critical polymer crystal nucleus is, on average, an anisotropic entity possessing a long axis (L), a median axis (M), and a short axis (S). The eigenvalues were used to assign these labels to the eigenvectors, and thus establish a local LMS frame of reference for each cluster. For reference, the center of mass (COM) of a nucleus corresponds to the origin of the local LMS frame of reference for that nucleus. In turn, property distributions where extracted for each nucleus using the LMS frame of reference, and these distributions were averaged together for nuclei corresponding to 240-360 carbon atoms inclusive to create average property distributions around pre-critical nuclei. The average distributions presented in this paper are based on only those snapshots where the largest nucleus in the corresponding system corresponded to 360 or fewer carbon atoms. This was done to ensure that the distributions reflect the initial nucleus-melt interface and not more complex, subsequent structures; multiple large-scale, post-critical clusters formed in some of the simulations (see Fig.~\ref{fig:potEner}C).
 
In this study, we consider two classes of distributions, radial and cylindrical. Each radial distribution captures the evolution of a local property (\eg density) as a function of radial distance ($r$) from the nucleus COM (see~Fig.~\ref{fig:schematic}), and is thus a 1D projection of the full 3D LMS distribution. A cylindrical distribution captures the evolution of a local property as a function of: 1) distance along the long axis of a nucleus from its COM ($z$), and 2) radial distance from the nucleus long axis ($r'$). Such a cylindrical distribution corresponds to a 2D projection of the full 3D LMS distribution. Radial and cylindrical distributions enabled improved sampling compared to the full 3D distributions. Radial distributions are generally presented in this study in order to facilitate comparisons between different properties.

\subsection{Crystalline Reference Data}
In order to compare the properties of nuclei and their surroundings to those of crystalline polyethylene, a second set of simulations was performed on a perfect polyethylene crystal. The methodological details of these simulations (\eg thermostating and barostating) were the same as those of the crystallization simulations at 285 K unless otherwise noted below. A perfect crystal consisting of 450 polyethylene chains was constructed according the crystal structure of long-chain normal hydrocarbons.\cite{Bunn1939} Each chain consisted of 240 coarse-grain beads (720 carbon atoms) and was connected to itself across the periodic boundaries of the simulation cell, effectively yielding an infinite polymer chain. The system thus lacked crystal defects arising from chain ends and folds. The perfect crystal was heated from 1 K to 250 K over 20,000,000 time steps ($\sim$407 ns). The temperature ramp was achieved by linearly increasing the set point of the simulation's Nos{\'e}-Hoover \cite{Nose1984,Hoover1985} chain thermostat\cite{MartynaJCP1992} using the internal functionality of the fix NPT command in LAMMPS. The crystal was then equilibrated at 285 K for 2,000,000 time steps ($\sim$20 ns).\footnote{This simulation used a time step that was half the length of the time step used in the other simulations.} A production simulation was then performed at 285 K for 1,000,000 time steps ($\sim$20 ns) to sample the properties of the equilibrated crystalline system. As with the nucleation simulations, snapshots were saved every 20,000 time steps ($\sim$0.4 ns). Local properties (\eg $P_{2}$) were calculated for all polymer segments in the system, and then averaged across all segments and snapshots to obtain average local property values for the perfect crystal.

\section{Results and Discussion}
\begin{figure}[!h]
\centering
\includegraphics[width=0.6\columnwidth]{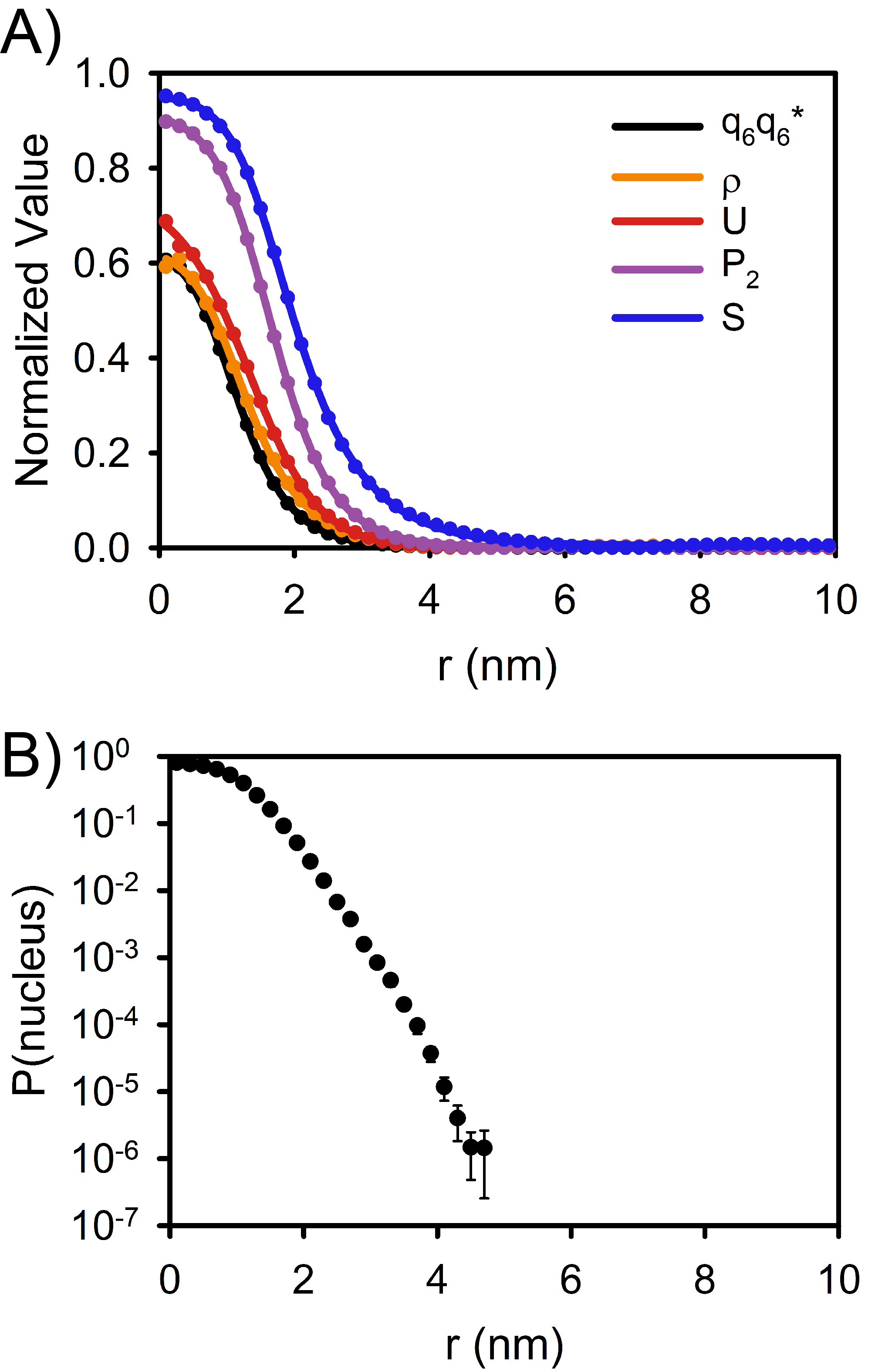}
\caption{Average variation in segment properties at increasing radial distances from nucleus COM for nuclei corresponding to 240-360 carbon atoms. A) Normalized property profiles. The profiles have been normalized so that values of 0.0 and 1.0 indicate property values corresponding those of the surrounding metastable \PE melt, and those of a perfect polyethylene crystal, respectively. The curves correspond to 5-parameter sigmoidal fits of the underlying data points ($R^2$ $>$ 0.99). The standard error range associated with each data point is much smaller than its symbol size except for the first points of $\rho$ and U at r =0.1 nm where the standard error ranges are comparable to the symbol sizes. B) The average probability of finding segments that are part of the nucleus at $r$=0 nm, $P(nucleus)$, as a function of increasing distance from that nucleus. Note that $P(nucleus)$ is shown using a log axis.}
\label{fig:relInterface}
\end{figure}

Normalized radial property distributions reveal that there is a partial spatial decoupling of segment properties at the nucleus-melt interface (see~Fig.~\ref{fig:relInterface}). For example, on approaching the center of a nascent nucleus (\ie r = 0 nm in Fig.~\ref{fig:relInterface}), surrounding polymer chains begin to display increasing alignment with the nematic director of the nucleus at  potentially up to $\sim$6 nm away from the nucleus as can be seen from the $S$ curve in Fig.~\ref{fig:relInterface}A. In contrast, 
changes in local density, local potential energy, and $q_{6}q_{6}^{*}$ begin to manifest only at much shorter distances in the vicinity of $\sim$2-3 nm. There is thus a spatial lag in some property transitions. As such, polymer segments approaching the crystal-melt interface experience increased order (reductions in their entropy) as evidenced by both $S$ and $P_2$ before achieving the lower enthalpy (potential energy) of the crystalline phase. This phenomenology is consistent with microscopic explanations of interfacial free energies in which the spatial separation of entropy losses and ethalpic payback makes it unfavorable to expand an interface.
\clearpage
\clearpage
Interestingly, the $q_{6}q_{6}^{*}$ order parameter fails to capture this expected interfacial phenomenology. In fact, the $q_{6}q_{6}^{*}$ data are nearly coincident with the density data in Fig.~\ref{fig:relInterface}A. This likely stems from two factors. First, $q_{6}q_{6}^{*}$ is based on a sum~\textemdash~not an average~\textemdash~over neighbouring beads within a spatial cutoff (see Eq.~\ref{eq:q6q6}), so $q_{6}q_{6}^{*}$ is sensitive to density changes in addition to orientational ordering. Second and as previously mentioned,  $q_{6}q_{6}^{*}$ depends on a much larger region of space than $P_2$, so it is likely not as sensitive to local ordering that may occur in interfacial regions. It is also important to note that all segments within 0.635 nm of a segment contribute equally to its $q_{6m}$ and $q_{6}q_{6}^{*}$ values while such segments do not contribute equally to a segment's potential energy; a segment that is farther away makes a smaller contribution in accordance with the spatial dependency of non-bonded interactions. Consequently, the $q_{6}q_{6}^{*}$ metric potentially possesses greater environmental sensitivity than potential energy, which explains why the $q_{6}q_{6}^{*}$ profile in Fig.~\ref{fig:relInterface} is displaced to the left of the potential energy profile. 

While the centers of pre-critical nuclei are close the crystalline state in terms of segment alignment (\ie $S$ and $P_2$ values), they still markedly differ from the crystalline state in terms of density, potential energy, and $q_{6}q_{6}^{*}$ (see Fig.~\ref{fig:relInterface}A). In fact, the normalized density and potential energy curves in Fig.~\ref{fig:relInterface}A exhibit values of approximately 0.6-0.7 at r = 0 nm, indicating that these properties have only transitioned $\sim$60-70\% of the way to the crystalline state. Nevertheless, the pre-critical nuclei used to construct Fig.~ \ref{fig:relInterface}A are still relatively large with respect to the critical nucleus size (\ie 240-320 carbon atoms vs. $\sim$600 carbon atoms\cite{HallShapeJCP2019}). For the conditions considered in this study, polymer nuclei do not exhibit a ``crystalline'' core surrounded by a crystal-melt interface as envisioned, for example, in classical nucleation theory. Rather, polymer nuclei exhibit properties intermediate between those of the melt and the crystal. Previous studies\cite{HallPolymers2020,HallShapeJCP2019} probing other facets of polymer nucleation similarly indicate that nuclei are not simply miniature versions of lamellar polymer crystals.

\begin{figure}[!ht]
\centering
\includegraphics[width=0.5\columnwidth]{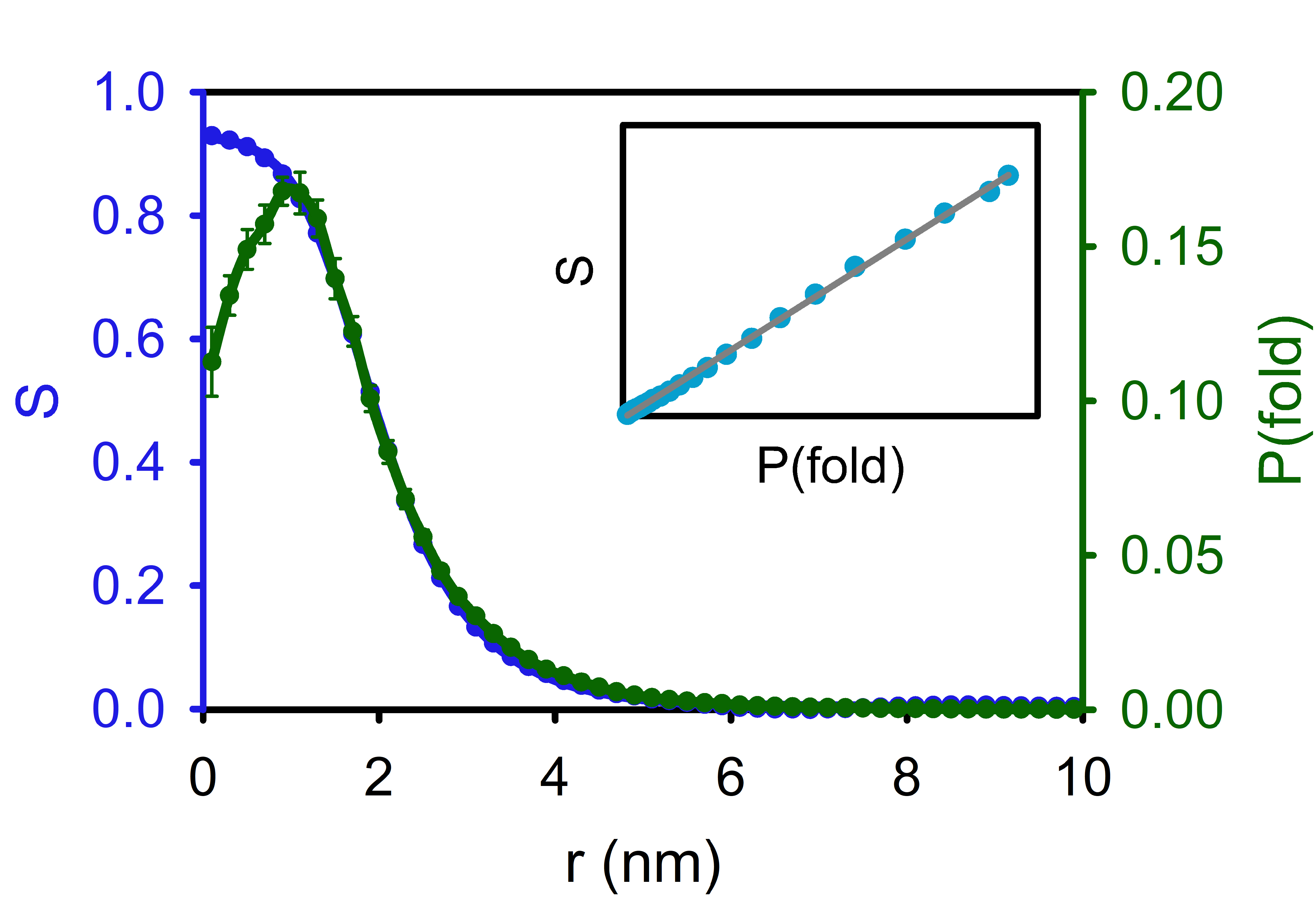}
\caption{Nematic alignment-fold correlation. The spatial evolution of average nucleus-segment alignment is overlaid with that of the average probability of encountering a segment that is part of a nucleus's folds, $P(fold)$, for nuclei corresponding to 240-360 carbon atoms. For reference and in accordance with our previous work,\cite{HallShapeJCP2019,HallModelJCP2019,HallHeatJCP2019} a segment was considered to be part of a fold if it was part of a set of noncrystalline segments (\ie $P_{2} < 0.85$) that was bordered by crystalline segments that were part of the same nucleus. The axes and curves are color-coded, and the error bars indicate standard errors. The inset provides the correlation between $P(fold)$ and $S$ for r=1.1-5.9 nm, and the line is a linear fit of the data ($R^2 > 0.99$).}
\label{fig:nematic-fold}
\end{figure}

Another notable feature in Fig.~\ref{fig:relInterface}A is the length scales over which the profiles deviate from the properties of the metastable melt (\ie values of 0), which is up to $\sim$6 nm in the case of nucleus-segment alignment ($S$). Importantly, nucleus-segment alignment is not simply a manifestation of the spatial extent of the polymer segments that are part of a nucleus (see Fig.~\ref{fig:relInterface}B). In particular, the probability of encountering a polymer segment that is part of the nucleus has already dropped below 0.01 by $\sim$2.4 nm, and it is several orders of magnitude lower at distances greater than 4 nm (Fig.~\ref{fig:relInterface}B). Therefore, the nematic alignment of a nucleus does not propagate from the nucleus into the surrounding metastable melt only through the polymer segments comprising the nucleus and their immediate neighbours (\ie contact pairs). Rather, nucleus-segment alignment propagation at the nucleus-melt interface is related to a nucleus's folds as evidenced by Fig.~\ref{fig:nematic-fold}. This relationship between the spatial extent of folds and nucleus-segment alignment is physically reasonable since folds are bounded by polymer segments that are part of the nucleus, and so propagation of nucleus alignment along a fold constitutes an intramolecular process. Moreover, previous work\cite{WelchJCP2017} on polymer melts has connected polymer crystallization to the development of on-chain (intramolecular) order, though the microscopic nature of this connection was not elucidated. Long-range intramolecular processes are relevant to polymer nucleus-melt interfacial phenomena, and relatively large-scale MD simulations are likely required to probe polymer crystallization phenomena.

Given that nucleus-segment alignment extends far beyond the region of space where there is a high probability of encountering the constituent segments comprising the nucleus, the full spatial extent of polymer structuring at polyethylene growth fronts and their interfaces may have been historically underestimated in some \textit{in silico} studies. More specifically, the length scales in Fig.~\ref{fig:relInterface}A may appear comparable to: 1) the $\sim$4-nm thicknesses revealed in Yamamoto's work\cite{YamamotoJCP2013,YamamotoJCP2008} on the tapered growth fronts of growing polyethylene crystals (see Fig. 9 in ref.~\citenum{YamamotoJCP2013} and Fig. 4 in ref.~\citenum{YamamotoJCP2008}), and 2) variations in crystallinity profiles during the growth of \textit{n-}C\textsubscript{50}H\textsubscript{102} crystals on polyethylene and tetrahedral substrates.\cite{BourqueEtAlJPCB2017,BourqueMacro2016} However, these studies probed growth fronts in terms of stem lengths and crystalline beads whereas the profiles in Fig.~\ref{fig:relInterface}A capture the spatial variation of properties in the vicinity of a nucleus arising from both constituent and non-constituent polymer segments. Profiles in the aforementioned studies likely do not fully capture structuring in the vicinity of melt-crystal interfaces. 

\begin{figure}[!t]
\centering
\includegraphics[width=0.5\columnwidth]{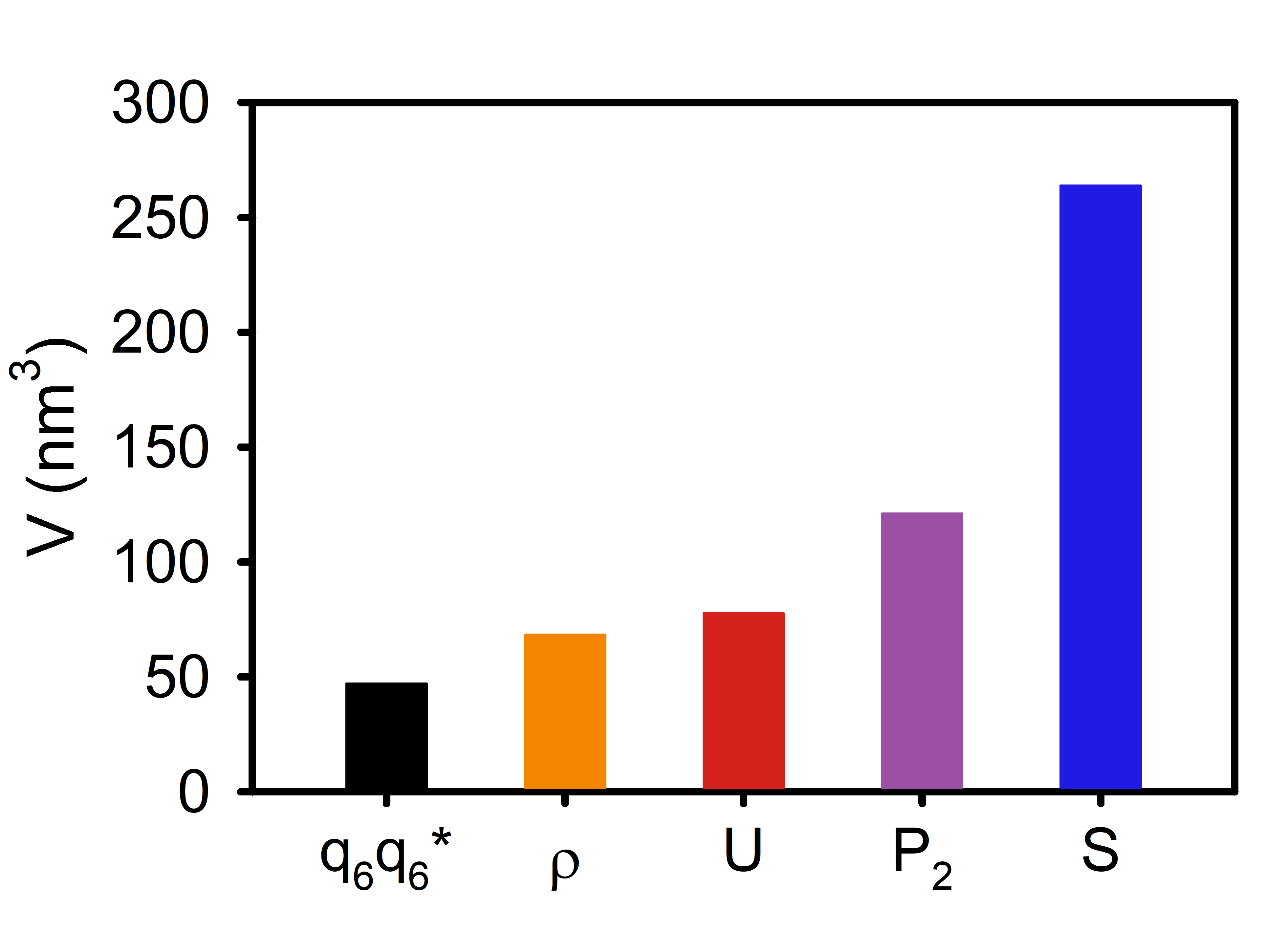}
\caption{Estimates of the average combined volume of a nucleus and its interface based different properties. Each property's volume estimate was constructed using the distance at which the corresponding radial profile in Fig~\ref{fig:relInterface} reached a normalized value of 0.05 (\ie 5\% of the transition from the melt to crystalline state) while invoking spherical geometry.}
\label{fig:volume}
\end{figure}

It is important to distinguish between long-range nucleus-segment alignment and local segment-segment alignment (\ie $S$ and $P_2$). In particular, the $P_2$ order parameter does not start to deviate from melt-like behaviour until one is less than $\sim$4 nm from a nucleus (Fig.~\ref{fig:relInterface}A). Therefore, polymer segments do not to exhibit enhanced alignment with their neighbouring segments until they are within $\sim$4 nm of a nucleus (with respect to the center of mass of that nucleus). In contrast, polymer chain segments start to align with the nematic director (\ie chain direction) of a nucleus at a distance $\sim$6 nm as demonstrated by the $S$ profile in Fig.~\ref{fig:relInterface}A. Long-range alignment and local packing are distinct, consistent with a nematic-like transition. These results thus indicate that a polymer nucleus can be considered to reside in a nematic-like droplet. In accord with this perspective, $S$-based estimates of the combined volume of a nucleus and its interfacial region are much larger than volumes obtained using other metrics, such as density and $P_2$, as can be seen in Fig.~\ref{fig:volume}. On approaching a nucleus, first long-range nematic alignment (S) increases, then local segment-segment alignment ($P_2$) increases, and then potential energy, density and $q_{6}q_{6}^{*}$ start to change (Fig.~\ref{fig:relInterface}A and Fig.~\ref{fig:volume}). As a result, polymer crystal nucleation involves a partial spatial decoupling of polymer properties for the quiescent, non-flow conditions considered in this study.

Our insights align well with results from previous computational studies\cite{ZhangLarsonJCP2019,TangetalJCP2018,TangPRM2017,AnwarEtAlJCP2014,AnwarJCP2013,LuoSommerMacro2011} on polymer and alkane crystallization. Zhang and Larson\cite{ZhangLarsonJCP2019} recently demonstrated that quenching polyethylene melts below their isotropic-nematic transition temperature ($T_{IN}$) results in rapid crystal nucleation on timescales $<$100 ns, which is much shorter than the microsecond induction times associated with this study's simulations (see Fig.~\ref{fig:potEner}).
To test whether our systems were susceptible to an isotropic-nematic transition prior to crystallization, average system-wide nematic order was calculated for the two systems that did not nucleate during their $\sim$4-$\mu$s simulations (\ie the systems lacking substantial potential energy dips in Fig.~\ref{fig:potEner}) using a procedure similar to what was used to establish the nematic director of individual nuclei. More specifically, the tensor order parameter $Q_{system}$ was constructed for each snapshot based on the backbone vectors of all polymer segments in the system 
in accordance with the work of Eppange and Frenkel.\cite{EppangeMolPhys1984} The nematic order of the system was taken to be the largest eigenvalue ($\lambda$) of $Q_{system}$; $\lambda$ approaches 1 with increasing alignment and zero for a bulk isotropic phase. This study's perfect crystal system yielded $\lambda$=0.976 $\pm$ 0.0002 (average $\pm$ standard deviation) while the average $\lambda$ value for the non-nucleating simulations is 0.014 $\pm$ 0.006. These results indicate that the supercooled \PE melts considered in this study generally correspond to isotropic states prior to nucleation and not a metastable nematic states. The non-zero $\lambda$ for the non-nucleating simulations is likely due to  transient pre-critical nuclei, which do exhibit chain alignment (nematic order). Consequently, this study considers polymer crystal nucleation under different conditions to Zhang and Larson.\cite{ZhangLarsonJCP2019} Nevertheless, our results can be used to interpret findings from the aforementioned study. In particular, Zhang and Larson\cite{ZhangLarsonJCP2019} concluded that rapid nucleation at $T<T_{IN}$ results from the formation of metastable nematic precursors that temporally precede and enhance crystal nucleation. Note that their work\cite{ZhangLarsonJCP2019} probed nematic order in terms of local segment-segment alignment (\ie using a variant of the $P_2$ order parameter) rather than probing long-range nucleus-segment order. It was also found that the onset of local segment-segment nematic alignment and crystallinity become temporally coincident as $T_{IN}$ is approached. In essence, previous work\cite{ZhangLarsonJCP2019} demonstrates that the development of local segment-segment alignment and crystallinity can become, at least partially, temporally decoupled under certain circumstances, but remain temporally coincident at higher temperatures. Our results demonstrate that spatial decoupling is present at the nucleus-melt interface. In turn, given that non-flow nuclei reside in droplets of long-range nematic ordering and local segment-segment order, processes that induce nematic order should enhance crystallization, such as the specific conditions considered in ref.~\citenum{ZhangLarsonJCP2019}. 

Based on fully atomistic simulations, Tang et al.\cite{TangPRM2017} found that hexagonal structuring precedes the formation of nuclei with orthorhombic structuring (\ie the preferred polyethylene crystal structure under ambient conditions) such that orthorhombic nuclei form inside droplets exhibiting nascent hexagonal ordering and through their coalescence. They also observed nascent hexagonal structuring temporally preceding changes in the Voronio volume (local density) of carbon atoms. Moreover, subsequent work by Tang et al.\cite{TangetalJCP2018} on shear-induced crystallization suggests that chain alignment and densification precede crystallization under shear conditions. For reference, Tang et al. assigned particles as crystalline using a metric based on spherical harmonics and similar in spirit to $q_{6}q_{6}^{*}$, so their observations could correspond to a scenario where the $q_{6}q_{6}^{*}$ and $\rho$ curves in Fig.~\ref{fig:relInterface}A are spatially (and thus temporally) distinct as a result of flow. While this study's coarse-grain simulations cannot be used to probe the interplay between hexagonal and orthorhombic structuring, the temporally staged nature of local ordering and densification is consistent with the spatial decoupling in Fig.~\ref{fig:relInterface}A. Computational work by Anwar et al.\cite{AnwarJCP2013,AnwarEtAlJCP2014} revealed that crystal nucleation in \textit{n-}C\textsubscript{20}H\textsubscript{42} and \textit{n-}C\textsubscript{150}H\textsubscript{302} melts  proceeds temporally via: 1) development of nematic alignment, 2) densification and, 3) establishment of local orientational order (\ie what they call crystallinity and estimated using a metric based on the Steinhardt $q_{6m}$ metric). Anwar et al. may have observed a more exaggerated difference between density and local Steinhardt-based order compared to the current study due to differences between: 1) the $\bar{q}_{6}$ metric used in their study and the $q_{6}q_{6}^{*}$ metric used in the current study, and 2) their decision to monitor the time evolution of segment properties by tracking only those segments comprising a nucleus at a particular point in time (\ie they did not track the evolution of the segments comprising a nucleus as that nucleus evolved). It is also worth noting that the exact nature of the temporal separations observed in their work depend on system conditions (\eg flow vs. quiescent). However, in general, the temporal progressions observed in the work of Anwar et al.\cite{AnwarJCP2013,AnwarEtAlJCP2014} are similar to the spatial progression observed as the center of a nucleus is approached (Fig.~\ref{fig:relInterface}A). Consequently, there is a correspondence between the temporal separation of changes in segment properties during polymer crystallization, and their spatial decoupling at the nucleus-melt interface.

Based on analysis of a single snapshot, Luo and Sommer\cite{LuoSommerMacro2011} found that there is a thin $\sim$0.8-nm region around the edge of a growing poly(vinyl alcohol) (PVA) lamellar crystal where PVA chains segments exhibit enhanced alignment while their local densities are still comparatively low. Their results thus support a spatial decoupling between local ordering and densification, consistent with the results in Fig.~\ref{fig:relInterface}A. Luo and Sommer's comparatively low estimate of the extent of spatial variations in segment properties could be due to: 1) their analysis being based on a single snapshot, which would be susceptible noise and have difficulty resolving the long tails in Fig.~\ref{fig:relInterface} and~\ref{fig:nematic-fold}, and 2) differences in the chemical nature of their PVA system and this study's \PE systems. Kumar et al.\cite{KumarMacro2017} probed the interfacial structure at the boundary between crystalline and amorphous lamellar domains in polyethylene. In the case of linear polyethylene, their interfacial width estimates based on density were smaller than those based on local $P_2$ values, which aligns with the density curve in Fig.~\ref{fig:relInterface}A being to the left of the $P_2$ curve. Though previous computational studies have not quantitatively probed the interfacial structural details of nascent polymer nuclei, the results of this study are consistent with insights from previous work.

The spatial decoupling of variations in polymer properties during crystal nucleation under quiescent, non-flow conditions (as considered in this study) also provides a lens for understanding previous work on flow-induced polymer crystallization.\cite{LiuTianMacro2014,ZhaoMacro2013,BalzanoPRL2008,GutierrezEtAlMacro2004,WangEtAlMacro2000,RyanFarad1999,SamonMacro1999,TerrillEtAlPolymer1998} For reference, flow-induced crystallization often, though not always, results in shish-kebab morphologies in which long fibrils of aligned, crystalline polymer segments are decorated with thinner crystalline lamellar lobes in a manner reminiscent to skewered food, hence their name. Previous experimental work\cite{SamonMacro1999} on the melt spinning of polyethylene and poly(vinylidene fluoride) revealed that 2D small-angle x-ray scattering (SAXS) patterns show structuring prior to the development of crystalline reflections in the 2D wide-angle x-ray scattering (WAXS) patterns. These results were interpreted as indicating that defective shish-like structures form first, yielding SAXS-level structuring without crystalline WAXS patterns. In turn, the shish structures mature and the kebabs develop, leading to crystalline WAXS reflections. These results for flow-induced crystallization can also be interpreted as consistent with the spatial decoupling  revealed in this study. For example, the defective shish precursor structures could correspond to large nematic droplets, such as the region of enhanced $S$ order in Fig.~\ref{fig:relInterface}A. As shown in Fig.~\ref{fig:relInterface}A and Fig.~\ref{fig:volume}, these nematic regions exhibit a large spatial extent well before their interior nuclei are fully crystalline (\eg in terms of $q_{6}q_{6}^{*}$), potentially supporting a lag between between the experimental detection of different types of ordering, especially if their spatial separation were enhanced.

Many other studies\cite{LiuTianMacro2014,ZhaoMacro2013,BalzanoPRL2008,WangEtAlMacro2000,RyanFarad1999,TerrillEtAlPolymer1998} on polymer crystallization under a variety of conditions have also noted the development of structure in SAXS patterns prior the detection of crystallinity by WAXS. Such results have been interpreted as evidence for the existence of: non-crystalline pre-ordered precursors,\cite{LiuTianMacro2014} conformationally distinct regions in metastable melts,\cite{RyanFarad1999} metastable precursors lacking crystallinity,\cite{BalzanoPRL2008} and ``large scale ordering prior to crystal growth.''\cite{RyanFarad1999} Similarly, based on spatial variations in SAXS and WAXS signals, Guti{\'{e}}rrez et al.\cite{GutierrezEtAlMacro2004} concluded that flow-induced crystallization involves the formation of quasi-ordered bundles of parallel polymer chains that serve as precursors to polymer crystallization. Consistent with the aforementioned experimental insights, our results for quiescent, non-flow nucleation show that segment alignment develops first as a polymer segment approaches a nucleus (or is consumed by the growing nucleus), and prior to changes in other properties such as density and $q_{6}q_{6}^{*}$. In essence, the decoupling of SAXS and WAXS signals can be viewed as a temporal manifestation of the spatial decoupling of segment properties at the nucleus-melt interface with crystallization conditions dictating the extent to which properties, and hence signals, are decoupled. Furthermore, based on the temporal separation of SAXS and WAXS patterns, Terrill et al.\cite{TerrillEtAlPolymer1998} concluded, ``The transformation from the disordered phase [parent melt] to the better ordered partially crystalline phase [resulting semi-crystalline state] proceeds continuously passing through a sequence of slightly more ordered states rather than building up a crystalline state instantaneously.'' Consistent with such a perspective, order develops gradually across the nucleus-melt interface (see Fig.~\ref{fig:relInterface}). Kanaya et al.\cite{KanayaMacro2013} experimentally observed that shearing polymer samples above their nominal melting temperatures can induce micron-scale shish-kebab precursors that are only 0.15\% crystalline. Still other experimental work\cite{SekiMacro2002} probing shear-induced crystallization in blends of long and short chains found that shear enhanced the formation of crystallization precursors, and connected the morphology of the precursors (\ie point-like vs. thread-like) to different types of flow-induced orientational order (\eg segmental vs. large-scale, long-chain orientation) as arising from different system conditions. These shear results also point to a decoupling of chain ordering and crystallization, and highlight that flow-related processes greatly enhance the nanoscale spatial decoupling present under quiescent non-flow conditions (Fig.~\ref{fig:relInterface}). Other work\cite{WangEtAlMacro2000} studying polymer crystallization with x-ray and polarized light (PL) scattering found that large domains possessing local orientational order precede crystallization, and it was proposed that domains of varying structure form in a temporally sequential, spatially-nested manner (\eg the PL-detectable domains appear first, and then the SAXS-detectable domains form afterwards in the PL-detectable domains). Such behavior is in accord with the spatial decoupling in the interfacial region of this study's non-flow nuclei in which segment alignment precedes other changes and property transitions are spatially nested (Fig.~\ref{fig:relInterface} and Fig.~\ref{fig:volume}). In summary, the results presented in this study for crystal nucleation under quiescent, non-flow conditions are consistent with those from experimental flow studies.\cite{LiuTianMacro2014,KanayaMacro2013,ZhaoMacro2013,BalzanoPRL2008,GutierrezEtAlMacro2004,SekiMacro2002,WangEtAlMacro2000,RyanFarad1999,SamonMacro1999,TerrillEtAlPolymer1998} In this paradigm,  observed phenomenology during the early stages of flow-induced polymer crystallization can be related to characteristic structural distinctions present in nuclei formed under quiescent, non-flow conditions. While there are many differences between crystallization under non-flow and flow conditions (\eg final morphologies), this work highlights one way in which there is conceptual continuity between the two.

\begin{figure}[!b]
\centering
\includegraphics[width=1.0\columnwidth]{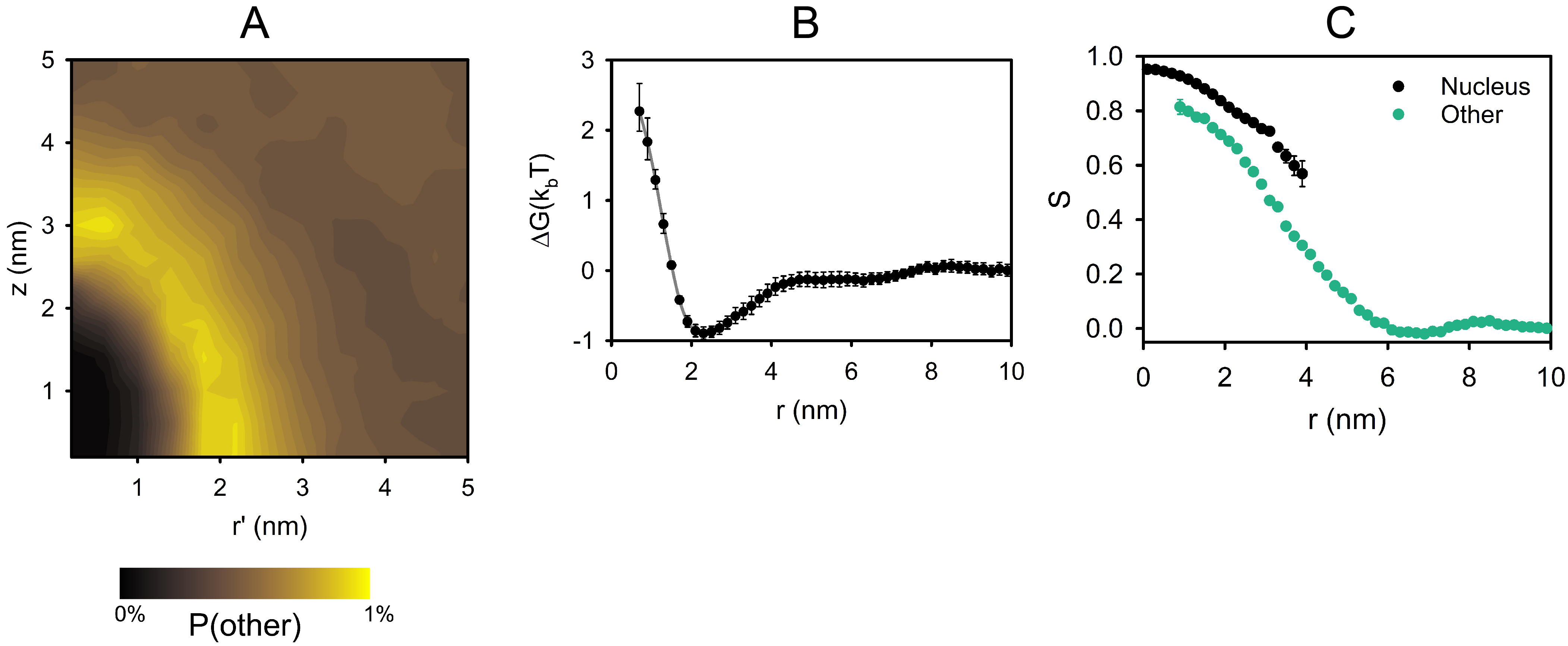}
\caption{The presence of other nuclei. A) The cylindrical distribution of the average probability of ordered polymer segments not part of a nucleus in the vicinity of that nucleus. These segments are thus part of the other nuclei, hence $P(other)$. Here, $z$ corresponds to the direction of the long axis of the nucleus while $r'$ is radial distance from the long axis of the nucleus. Note ($r',z$)=(0 nm, 0 nm) corresponds to the nucleus COM. B) The average relative free energy of ordered polymer chain segments not part of a nucleus at increasing distances from the nucleus (nucleus COM, $r$ = 0 nm). The profile has been estimated using $G=-k_{b}TlnP(Other)$, and then subtracting off the free energy minimum. C) Average radial profiles of nucleus-segment alignment for: 1) ordered polymer segments that are part of the nucleus with its COM at $r$=0 nm (black data), and 2) ordered polymer segments that are  part of other nuclei (green data). Points are only shown for distances where the radial averages were well sampled (\ie $n_{sample} \geq 30$). The error bars in Panels B and C correspond to standard errors. The distribution and profiles in Panels A-C are for a nucleus corresponding to 240-360 carbon atoms residing at $r$=0 nm and ($r',z$)=(0 nm, 0 nm).}
\label{fig:2dfig}
\end{figure}

\begin{figure}[!t]
\centering
\includegraphics[width=0.6\columnwidth]{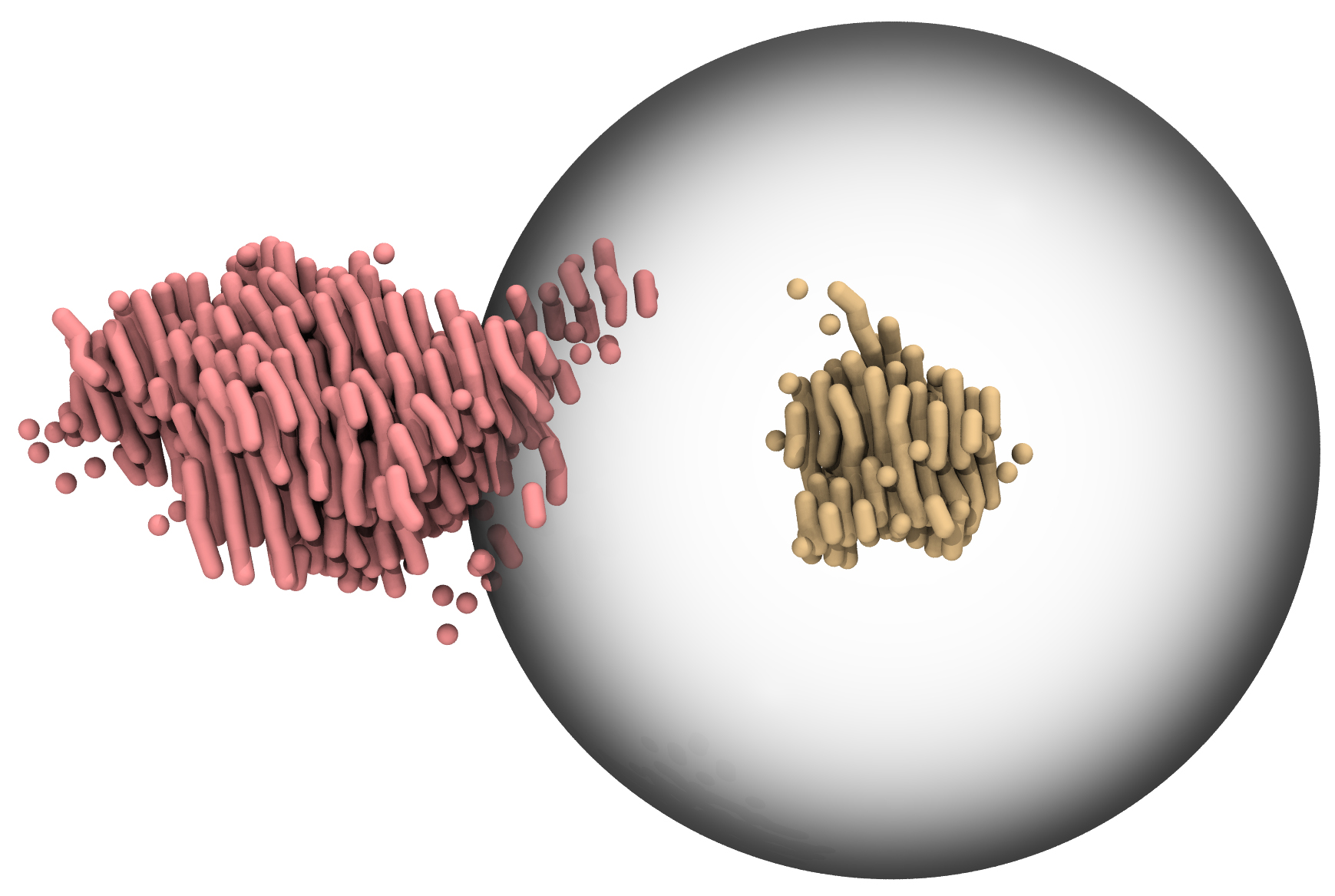}
\caption{Two nuclei from a snapshot at $\sim$1.8 $\mu$s in the simulation corresponding to the black curve in Fig.~\ref{fig:potEner}A. The black sphere has a radius of 6 nm, and is centered on the COM of the yellow nucleus. The red and yellow nuclei correspond to 2,703 and 597 carbon atoms, respectively.}
\label{fig:twoCluster}
\end{figure}

Given that nematic ordering and chain alignment enhance crystallization\cite{ZhangLarsonJCP2019,HuEtAlMacro2002} and that a nucleus resides in a nematic droplet, it is reasonable to expect that the formation of one nucleus could induce the formation other nuclei. To explore this theory, we calculated the average probability of finding ordered polymer segments ($P_{2} \geq 0.85$) as a function of distance from the center of each nucleus corresponding to 240-360 carbon atoms while excluding the constituent polymer segments of that nucleus. The resulting spatial distribution thus corresponds to the average probability of finding polymer segments that are part of other nuclei, $P(other)$, in the vicinity of a given nucleus. As can be seen in Fig.~\ref{fig:2dfig}A, there is on average a region around each nucleus where there is an enhanced probability of finding ordered segments that are not part of the nucleus. In fact, it is approximately $\sim$1 k\textsubscript{b}T more favourable for an ordered polymer segment to be in this band than in the surrounding
\noindent metastable melt (see Fig.~\ref{fig:2dfig}B). The non-constituent ordered segments in the vicinity of a nucleus are less aligned with the nematic director of the nucleus than those segments comprising the nucleus as shown in Fig.~\ref{fig:2dfig}C. Importantly, this difference in alignment behavior is statistically significant (paired-data single-tail t-test on region where the curves overlap in Fig.~\ref{fig:2dfig}C, level of significance: 0.05, calculated p-value: $4.33\times10^{-7}$). Therefore, the non-constituent ordered segments in the vicinity of a nucleus do differ from those comprising the nucleus. Moreover, the enhancement band is well within the nematic droplet of the nucleus (compare Fig.~\ref{fig:2dfig}A-B with the $\sim$6 nm extent of the $S$ profile in Fig.~\ref{fig:relInterface}A), providing strong evidence that the nematic droplet of one nucleus can serve as the birthplace of other nuclei. Note that just because one nucleus can enhance the formation of other nuclei, it does not mean that the resulting nuclei will overcome the free energy barrier to crystallization, and become distinct, post-critical entities. For example, a nucleus might induce the formation of surrounding nuclei, and these induced nuclei might subsequently coalesce with the original nucleus as it expands, contributing to polymer crystal growth. In this vein, the two nuclei in Fig.~\ref{fig:twoCluster} are spatially close as highlighted by the 6-nm black sphere centered on the yellow nucleus, and are experiencing each other's alignment field based on the length scales in Fig.~\ref{fig:relInterface}A. As the simulation progresses, the red nucleus grows laterally and consumes the region of space around the yellow nucleus, eventually resulting in the purple crystallite in Fig.~\ref{fig:potEner}. This coalescence is likely facilitated by the alignment between the nuclei; the angle between the nematic directors of the two nuclei is 17.6\textdegree, which yields $S$=0.86 when substituted into Eq.~\ref{eq:S}. Nevertheless, the formation of one nucleus could initiate the formation of subsequent distinct nuclei (\ie crystalline clusters) through a cascade-like process, consuming the metastable melt and transforming it to a semi-crystalline solid state. Consistent with such a perspective, previous experimental work on shear-mediated polymer crystallization\cite{SekiMacro2002} has proposed that thread-like morphologies form through the attachment of long polymer chains to a nascent point-like nucleus, and that the dangling segments of the chains then become oriented in front and behind the point-like nucleus, leading to the enhanced formation of nuclei ahead of and behind the nucleus (along the shear direction) and ultimately thread formation. Moreover, subsequent \textit{in silico} work\cite{NieEtAlMacro2018} on shear-induced nucleation supports such a mechanism. A number of \textit{in silico} studies on polymer crystallization have also observed the formation of multiple, distinct crystallites in a single system (\eg see ref.~\citenum{ZhaiEtAlNano,ZhaiEtAlMacro2019,MoyassariEtAlMacro2019,TangetalJCP2018,TangPRM2017,MorthomasEtAlPhysRevE2017,LuoPolymer2017,LuoEtAlMacro2016,LuoSommerPRL2014,YamamotoJCP2013,YamamotoJCP2008}). Nucleus-induced nucleus formation is likely an important feature of both flow-induced and quiescent polymer crystallization, and relevant to the morphology of semi-crystalline polymer samples.

\section{Conclusions}
Through a detailed analysis of the structure of pre-critical nuclei, this study has demonstrated that the central regions of nascent nuclei exhibit far from crystalline behaviour in terms of density, potential energy, and $q_{6}q_{6}^{*}$ while they may be viewed as approaching crystalline in terms of chain alignment metrics (\eg $P_2$ and $S$). Consistent with these differences, the region of space over which polymer segments transition from melt-like to solid-like is metric dependent; there is a partial spatial decoupling of polymer properties. In particular, nucleus-segment alignment can be observed at distances up to $\sim$6 nm from a nucleus whereas variations in density only occur when polymer segments are much closer to the nucleus (i.e., within $\sim$3 nm of its COM). These results highlight that nuclei and their interfaces can extend over a substantial region of space even if they are pre-critical as considered in this study. The broad spatial extent of nucleus-segment alignment compared to other metric suggests that nuclei reside in nematic-like droplets even under quiescent non-flow conditions. Moreover, the extent of the nematic droplet is connected to the folds associated with a nucleus. Through the lenses of spatial decoupling and nematic droplets, we connected our findings to observations from previous experimental and computational studies on polymer crystallization under flow and non-flow conditions, thereby conceptually linking two polymer crystallization regimes that are often viewed as distinct. In particular, the temporal separation of SAXS and WAXS structuring during flow-induced crystallization can be viewed as a manifestation of flow selectively enhancing spatial property decoupling already present in non-flow nuclei. While this study has probed quiescent polyethylene crystal nucleation, it is reasonable to expect that spatial property decoupling at the nucleus-melt interface and associated phenomena are general features of the crystallization of semi-crystalline polymers.

This study also revealed that the nematic droplet around a nucleus corresponds to a region of space where there is enhanced formation of ordered polymer chain segments that are not part of the nucleus. It was demonstrated that these surrounding ordered segments are statistically distinct from those comprising the nucleus, highlighting the possibility of one nucleus inducing the formation of other nuclei. Such nucleus-induced nucleus formation is consistent conclusions from previous experimental work, and is relevant to the morphological development of semi-crystalline polymeric materials. The interfacial details revealed in this study provide a new lens for understanding structural and temporal evolution during the earliest stages of polymer crystallization.
\clearpage
\begin{acknowledgement}
This work was supported by the U.S. Army Research Laboratory through contracts W911NF-18-9-0269 and W911NF-16-2-0189. This study used the HPC facilities at Temple University, which were partially funded by the National Science Foundation through a major research instrumentation grant (grant number: 1625061). M.L.K. acknowledges the support of H.R.H. Sheikh Saud through a Sheikh Saqr Research Fellowship.
\end{acknowledgement}

\bibliography{bibliography}

\end{document}